\documentclass[sigconf]{acmart}

\usepackage{caption}
\usepackage{subcaption}
\usepackage{makecell}
\usepackage{bbding}

\AtBeginDocument{%
  \providecommand\BibTeX{{%
    \normalfont B\kern-0.5em{\scshape i\kern-0.25em b}\kern-0.8em\TeX}}}

\copyrightyear{2022} 
\acmYear{2022} 
\setcopyright{acmcopyright}
\acmConference[MM '22]{Proceedings of the 30th ACM International Conference on Multimedia}{October 10--14, 2022}{Lisboa, Portugal}
\acmBooktitle{Proceedings of the 30th ACM International Conference on Multimedia (MM '22), October 10--14, 2022, Lisboa, Portugal}
\acmPrice{15.00}
\acmDOI{10.1145/3503161.3548168}
\acmISBN{978-1-4503-9203-7/22/10}

\begin{document}

\title{Blind Robust Video Watermarking Based on Adaptive Region Selection and Channel Reference}

\author{Qinwei Chang}
\affiliation{%
  \institution{AI Technology Center, Tencent}
  \city{Beijing}
  \country{China}
}
\email{wadechang@tencent.com}

\author{Leichao Huang}
\affiliation{%
  \institution{AI Technology Center, Tencent}
  \city{Shenzhen}
  \country{China}
}
\email{darielhuang@tencent.com}

\author{Shaoteng Liu}
\affiliation{%
  \institution{AI Technology Center, Tencent}
  \city{Beijing}
  \country{China}
}
\email{liushaoteng@live.com}

\author{Hualuo Liu}
\affiliation{%
  \institution{AI Technology Center, Tencent}
  \city{Shenzhen}
  \country{China}
}
\email{karlindliu@tencent.com}

\author{Tianshu Yang}
\affiliation{%
  \institution{AI Technology Center, Tencent}
  \city{Beijing}
  \country{China}
}
\email{gracetsyang@tencent.com}

\author{Yexin Wang}
\affiliation{%
  \institution{AI Technology Center, Tencent}
  \city{Shenzhen}
  \country{China}
}
\email{yexinwang@tencent.com}

\begin{abstract}

Digital watermarking technology has a wide range of applications in video distribution and copyright protection due to its excellent invisibility and convenient traceability. This paper proposes a robust blind watermarking algorithm using adaptive region selection and channel reference. By designing a combinatorial selection algorithm using texture information and feature points, the method realizes automatically selecting stable blocks which can avoid being destroyed during video encoding and complex attacks. In addition, considering human's insensitivity to some specific color components, a channel-referenced watermark embedding method is designed for less impact on video quality. Moreover, compared with other methods' embedding watermark only at low frequencies, our method tends to modify low-frequency coefficients close to mid frequencies, further ensuring stable retention of the watermark information in the video encoding process. Experimental results show that the proposed method achieves excellent video quality and high robustness against geometric attacks, compression, transcoding and camcorder recordings attacks.

\end{abstract}




\renewcommand{\shortauthors}{Qinwei Chang et al.}
\maketitle

\section{Introduction}
With competition intensifying in video streaming, content owners and providers need to be on top of their security measures more than ever before to protect against content piracy. Video digital watermarking solutions can play an important role in deterring and reducing digital piracy. It is mainly used for video content protection and transaction tracking to detect illegal use, modification, and distribution of the content. 
\par

In order to support tracking, a watermarking scheme has to fulfil specific requirements. First, it is necessary to ensure that the video quality is not affected when embedding the watermark, and it must be imperceptible to human observers. Next, the watermarking scheme has to be robust enough. Videos usually go through multiple distributions by pirated. The pirated videos are often re-encoded and compressed during the transmission process. Therefore, video watermarking must be robust against geometric, compression and camcorder recordings attacks. Last but not least is the processing time which has been an essential criterion for assessing and selecting the optimal technique. 
\par

The process of inserting a watermark into a video is usually called embedding, and the process of getting watermark information from a watermarked video is called extraction. The need of the original data during extraction categorizes watermarking schemes as the non-blind algorithm. Watermark extraction of the non-blind algorithm is usually easier and more robust. However, in the case of most practical applications, the original host video is not available for watermark extraction. Therefore, the watermark should be detectable without reference to the original video content, i.e., the extraction should be blind. 
\par
In this paper, we propose a novel blind video watermarking scheme to improve the robustness further. First of all, we use texture factor to select the appropriate regions to embed the watermark, which can avoid selecting the block with a larger degree of compression. What's more, we also introduce ORB key points to select regions against geometric attacks and camcorder recording attacks. ORB feature points have better stability to translation, rotation, and noise, which can avoid the loss of the watermark information during multiple complex attacks. The watermark embedding is performed by modifying the a low frequency band slightly close to the intermediate frequency since the high-frequency coefficients are not stable in video encoding and the low-frequency coefficients will affect the image quality.
\par

The main contributions of our proposed scheme are summarized as follows:

(1) We design a adaptive region selection method based on texture information and feature points, which reduces the impact on image quality and ensures the detection rate.
\par
(2) Considering human's insensitivity to specific color components, we propose a channel-referenced watermark embedding strategy by modifying low-mid frequencies to ensure higher robustness and less image quality loss.
\par


The remainder of this paper is structured as follows: in Section 2, related work is discussed. Next, our proposed watermarking scheme is described in details in Section 3. After the presentation of evaluation results in Section 4, the last section concludes the paper and future works.

\section{Related work}
Extensively research has been carried out for digital video watermarking systems, and numerous solutions have been proposed. Existing schemes for video watermarking can be mainly classified into three different domains: spatial-based, frequency-based and compress-based schemes. Each category is described individually as below:
\par

\textbf{Spatial-based.} This approach tries to modify the pixel values of the original image  to achieve the purpose of embedding watermark information. The main spatial domain methods include least significant bit (LSB) modification\cite{jindal2016performance}, spread spectrum modulation\cite{cox1997secure}, and so on. H. Kaur and E. V. Kaur\cite{kaur2016invisible} proposed an invisible video watermarking algorithm using an optimized LSB technique.To improve the robustness, Bayoudh et al.\cite{bayoudh2018online} proposed a multi-sprites dynamic video watermarking algorithm based on speed-up robust features (SURF), which can effectively resist collusion and transcoding attacks.
\par

\textbf{Frequency-based.} This approach tries to embed watermarks by performing modifications in the frequency domain. To prevent high-definition (HD) videos from unauthorized copying, Cheng et al.\cite{cheng2013recoverable} proposed a recoverable video watermarking algorithm in DCT domain based on code division multiple access (CDMA) modulation.To prevent the collusion attack, Gupta et al.\cite{gupta2018efficient} adopted DWT to resize frames into 512 × 512 based on security model, and the maximum mean values of LL and HL bands were used to select watermark positions. Depending on the energy compression property of DCT and the multiresolution property of DWT, some hybrid DWT/DCT-based video watermarking algorithms have been proposed \cite{joshi2017combined,kunhu2016index}.
\par

\textbf{Compress-based.} The video watermarking algorithm after compression searches for redundant space in the  compressed bit stream and embeds watermark information into it. To resist scaling attacks, Wang and Pearmain\cite{wang2006blind1} proposed a MPEG-2 video watermarking algorithm based on shadow-frame generation in the compressed domain combined with DCT transform. Based on the newly proposed bit stream syntax elements in H.264 standard, Li et al.\cite{li2012reference} embedded a watermark into the index of the reference frame during video encoding.  Aiming at the HEVC coding process, Yang and Li\cite{yang2018efficient}  proposed an efficient information hiding algorithm based on motion vector space coding.
\par

In addition to watermarking algorithms in different domains, a few studies have dealt with digital watermarking against camcorder recording attacks. Wang et al.\cite{wang2006blind2} proposed two blind MPEG-2 video watermarking methods: one can deal with horizontal cropping and the other down-scaling. Celik et al.\cite{celik2007camcorder} proposed a watermarking method for MPEG-2 videos that can survive camcorder recording. It can reduce watermark embedding complexity and minimize bit-rate increase by modulating only quantization matrices used for video encoding. Their technique, however, exhibits weakness to rotation and projection attacks. Matthew et al.\cite{tancik2020stegastamp} proposed a learned steganographic algorithm to enable robust encoding and decoding of arbitrary hyperlink bit strings into photos. The method requires numerous data to train deep neural networks, therefore incurring complexity and cost.

\section{Proposed method}
\begin{figure}[htbp]
\centering
\includegraphics[scale=0.28]{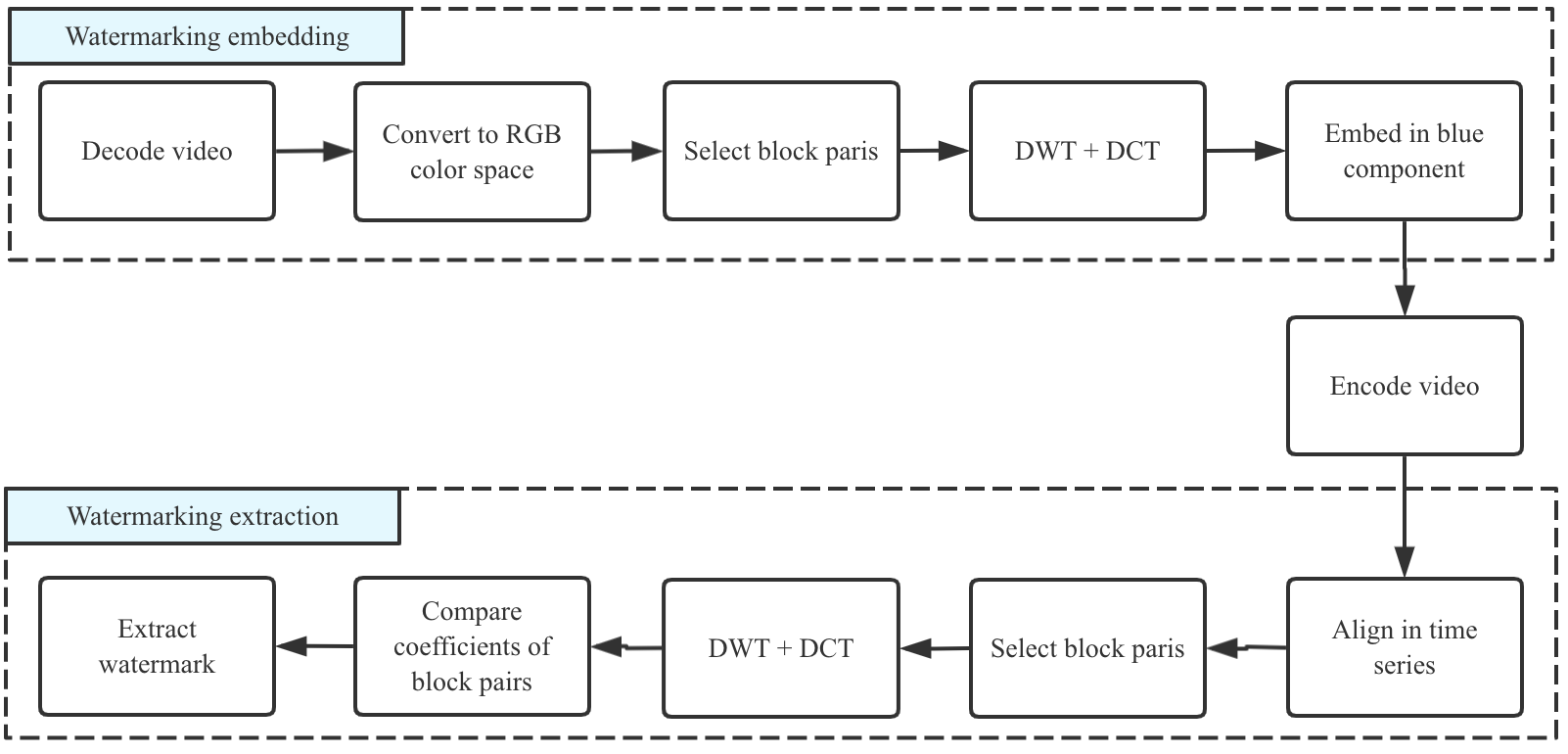}
\caption{The framework of watermarking embedding and watermarking extraction}
\label{fig:embedding and extraction}
\end{figure}

%
In this section, a robust watermarking method has been proposed against compression, geometric attacks and camcorder recording attacks. 
As show in Fig~\ref{fig:embedding and extraction}, by taking advantage of some stable feature points, the algorithm uses texture information of frames to embed watermarks in the frequency domain of the RGB color space.
And in extraction process, a blind watermarking extraction algorithm is introduced. 


\subsection{Selecting watermark embedding blocks}
\begin{figure}[t]
\centering
\includegraphics[scale=0.28]{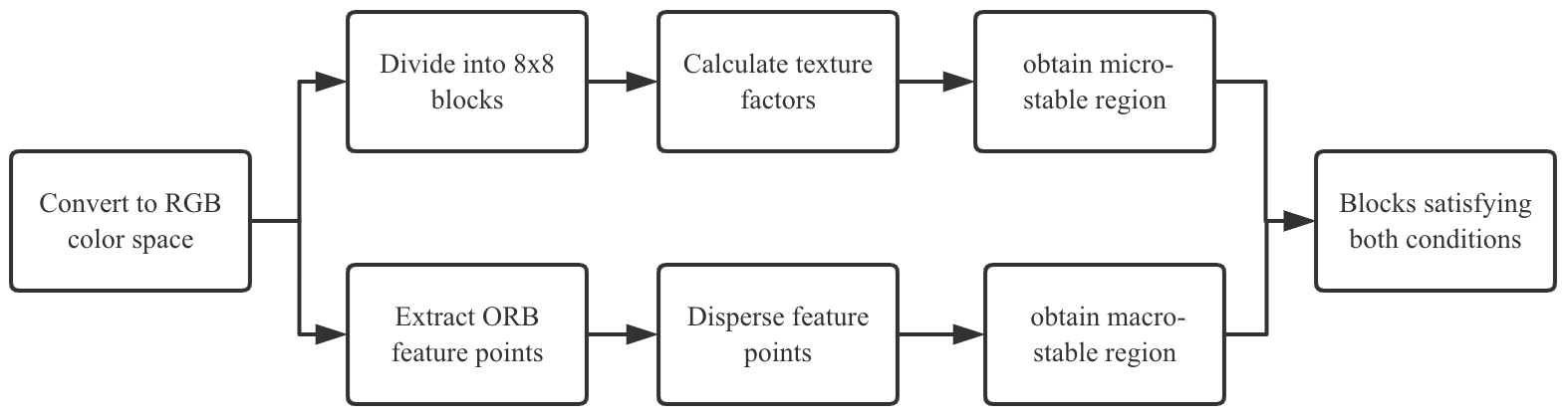}
\caption{Process of block selection. Texture factors are designed to select micro-stable regions that resist compression attacks. Feature points are designed to select macro-stable regions that resist geometric attacks and camcorder recordings.}
\label{fig:block selection}
\end{figure}

In the video encoding process, information of smooth area will be concentrated into low frequency coefficients while details will be centralized at intermediate and high frequency coefficients.
After that, most of the intermediate or high frequency coefficients are transformed and quantified to zero.
Therefore, when embedding the watermark, if the watermark information is embedded in the high frequency region, the quantization process will cause the loss of the watermark information during video coding. 
However, if the watermark is completely embedded in the low frequency area for the stability of the watermark, the image quality will be under negative influence.
Based on the above analysis, we designed a region selection method based on feature points and texture information as show in Fig~\ref{fig:block selection}, which not only reduces the impact on image quality but also ensures the detection rate.

\textbf{Texture factors.}
As mentioned in~\cite{ding2021compressed}, texture factors can use the content information in the video frame to select the appropriate block to embed the watermark. 
This design is in good agreement with the H.264 coding rules for the video compression process. 
In H.264 coding rules, the video frame is usually divided into 4x4 and 8x8 macroblocks, and the content of the macroblocks is quantized and compressed according to the motion information. 
Therefore, the block selected by using the texture factor can avoid choosing the block with a larger degree of compression as much as possible, and can better retain the watermark information.

The texture factor can be described as follows:
\begin{equation}
    \label{eq:texure factor}
    S(i) = N[Ef (i) + Rf (i) + Ef (i) \cdot Rf (i)]
\end{equation}
where i represents the $i^{th}$ $4 \times 4$ block, and N represents the normalization operation.
$Ef$ and $Rf$ can be calculated as:
\begin{equation}
    \label{eq:rf}
    Rf(i) = \|C\|_0
\end{equation}
\begin{equation}
    \label{eq:ef}
    Ef(i)=\sum_{j=1}^4{\sum_{k=1}^4{|c(j,k)|}}
\end{equation}
in which $\|C\|_0$ is the number of non-zero elements in DCT coefficient matrix of a $4\times4$ block and $c(j, k)$ represents the coefficient of the $j^{th}$ row and $k^{th}$ column of DCT coefficient matrix.

The texture features of visual sensitive areas are complex, and their $Rf$ is generally greater than the other areas.
Meanwhile, the area with more information still have more information after compression with different quantization parameters. 
And $Ef$ of these area is also larger than others. 

It should be noted that the above algorithm is based on the premise that all compression and quantization are conducted on $4\times4$ blocks. 
But in fact, $8\times8$ macroblocks are also involved in H.264 compression. 
Therefore, our algorithm improves the above process and designs a more general block selection method for H.264.
The block selecting rule can still be calculated with Equation 1, but $i$ now represents an $8\times8$ block instead of $4\times4$. 
For $Rf$, the improved calculation method is as follows:
\begin{equation}
    \label{eq:rfrf}
    Rf(i) = \sum_{m=1}^{4}{\|C(m)\|_0}
\end{equation}
where $m$ represents the index of $4\times4$ sub-blocks of a $8\times8$ block.
$C$ represents the DCT coefficient matrix of $m^{th}$ block.
As for $Ef$, the calculation is modified as follows:
\begin{equation}
    \label{eq:efef}
    Ef(i)=\sum_{m=1}^4{\sum_{j=1}^4{\sum_{k=1}^4{|c(j,k)|}}}
\end{equation}
where $c(j,k)$ represents the coefficient of the $j^{th}$ row and $k^{th}$ column of the $m^{th}$ DCT coefficient matrix.

It can be seen from Equation~\ref{eq:rfrf} and Equation~\ref{eq:efef} that we first divide the picture into $8\times8$ blocks in order to match the macroblock division of the encoding process as closely as possible. 
Then the selection of the block is placed in a $4\times4$ block. 
The voting of four $4\times4$ blocks determines whether to choose this $8\times8$ block.

\textbf{Feature point.}
Although the texture factor can help us choose a stable block to prevent the watermark from being damaged by compression attacks. 
However, when the video is subjected to geometric attacks and camcorder recording attacks, the information embedded in these blocks may not be well preserved. 
Therefore, in addition to the texture factor, we also introduce ORB keypoints for region selecting.

ORB feature points have better stability to translation, rotation and noise. 
After using the image pyramid, the ORB feature is scale invariant, and the robustness to attacks such as filtering and compression is further improved. 
Therefore, appropriate ORB feature points can be selected, and the block area can be constructed with these points as a reference.

From the properties of ORB feature points, it can be known that the larger the response value of the feature point is, the more robust the feature point is, and it is easier to retain under the geometric attack and the camcorder recording attack. 
Therefore, we select feature points with larger response values to select block regions.

\begin{figure}[htbp]
    \centering
    \includegraphics[scale=0.2]{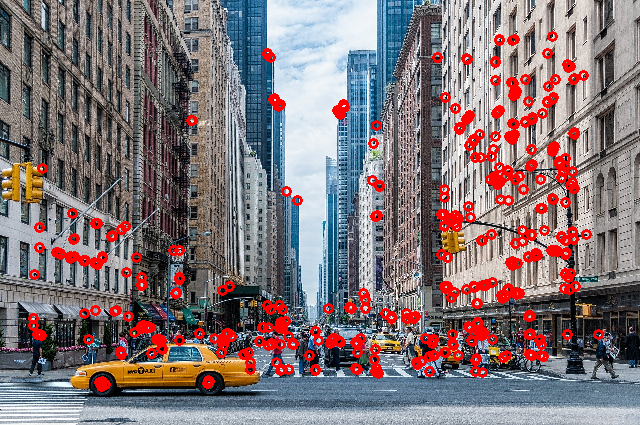}
    \includegraphics[scale=0.2]{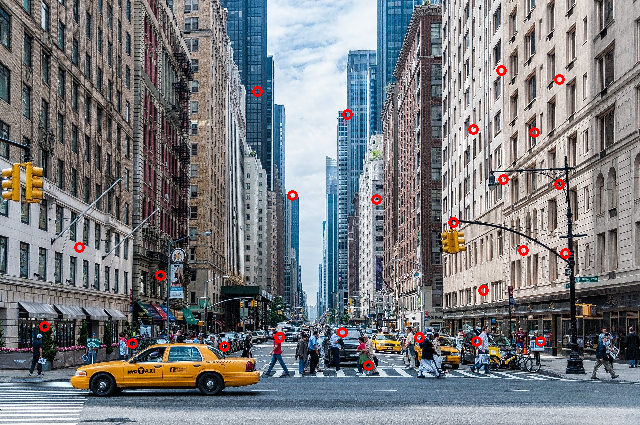}
    \caption{The figure above is the area represented by the direct calculation of ORB feature points. Obviously a large number of points are clustered on the same feature area. The bottom figure represents the ORB feature points after local clustering. It’s apparent that the number of feature points is significantly reduced, and there will only be one ORB feature point in a local region.}
    \label{fig:orb select}
\end{figure}

However, as shown in Fig~\ref{fig:orb select}, if the feature points with larger response values are directly selected, the phenomenon of feature point aggregation will occur. 
The clustering of feature points will cause all eligible blocks to be clustered in a small area, rather than scattered in the whole image. 
When encountering clipping or noise attack, the probability of watermark information loss will increase. 
Therefore, it is necessary to disperse the selected feature points on the whole image as much as possible. 
To solve this problem, we perform local clustering on the feature points of each frame, take each feature point as the center, and set the original radius as r. Only the feature points with the largest response value in the circle are retained, and other feature points are discarded. 
Here we set the radius to $4\sqrt{2}$ to ensure that each circle covers an $8\times8$ block.

From the analysis of the above two points, we can conclude that the texture factor tends to select stable blocks that are not destroyed by compression attacks from the microscopic level to de-embed watermarks. 
And the ORB feature points tend to analyze the stable area from a macro perspective to prevent the watermark area from being damaged by human attacks such as geometric attacks and video recorders. 
Combining the above two points, we can give the selection rules of the watermark area
\begin{equation}
    \label{eq:total B}
    B=S_{m=1} \cap OB
\end{equation}
Among them, $S_{m=1}$ represents the block sets of $S(i)=1$, and $OB$ represents the sets of block areas screened by ORB feature points. $B$ represents the screening area that complies with both rules at the same time, that is, the area where the watermark is embedded.

\subsection{Watermarking embedding}

After determining the area where the watermark is embedded, we need to imperceptibly embed the watermark into frames. 
%
The embedding process can be described as follows:

(1) Use the Arnold scrambling algorithm to preprocess the watermark sequence.
\begin{equation}
    W=\{w(i)|i=1,2,\dots L,w(i) \in {+1,-1}\}
\end{equation}

(2) Decode the video frame and transfer it to the RGB color space.

(3) Divide blocks according to Equation~\ref{eq:total B}.

(4) With the reference to Human Visual System (HVS), it is easily known that in the RGB color space, the human eye is the least sensitive to blue, so we consider modifying the blue component for watermark embedding. 
At the same time, in addition to the blue component to be modified, we also need to select another component as a reference for modification. 
Here we choose the green component as the reference component. DWT and DCT are performed after dividing the blue and green components into blocks.

(5) For each position selected from Equation~\ref{eq:total B}, we get DCT coefficient matrices of a pair of blocks named $BB$ and $BG$ ($BB$ represents the matrix of the blue component, $BG$ represents the green component matrix). 
Then we select a specific frequency band $f$ and calculate the sum of the coefficients of $BB$ and $BG$ in this frequency band respectively, denoted as $cB$ and $cG$. Next the difference $dB$ between the $cB$ and $cG$ is calculated:
\begin{equation}
    \label{eq:dif}
    dB=cB-cG
\end{equation}
In our work, the value of $f$ is 6.

(6) When the embedded information is 1, it is necessary to ensure that $cB$ is greater than $cG$. The modification of $cB$ can be described as follows:
\begin{equation}
    \label{eq:embedding +1}
    cB=\left\{
        \begin{aligned}
            &cB& {dB}>0 \\
            &cB + (-dB+p) &dB\leq0
        \end{aligned}
        \right.
\end{equation}
When the embedded information is -1, it is necessary to ensure that $cB$ is less than or equal to $cG$. $cG$ is modified as follows:
\begin{equation}
    \label{eq:embedding -1}
    cB=\left\{
        \begin{aligned}
        &cB& {dB}<0 \\
        &cB + (-dB-p) &dB\geq0
    \end{aligned}
    \right.
\end{equation}
where $p$ represents the embedding strength

(7) For the modified $cB$, calculate the average modification amount of each coefficient on the band
\begin{equation}
    \label{eq:average to each coefficient}
    aB=\frac{cB}{f}
\end{equation}
Then each coefficient on band $f$ is modified to $aB$.

(8) Conduct IDCT and IDWT on the modified block to get the frame with the watermark embedded.

It can be seen from the embedding process that we embed the watermark information by modifying the DCT coefficients of the middle and low frequency bands. 
And in order to reduce the impact on the image quality as much as possible, in this solution, the value of a certain low-frequency coefficient is not singly modified. 
Instead, the modified value is amortized over multiple coefficients in the same frequency band. 
In addition, considering the compression of image content by video coding, we adopt the method of embedding one bit information in consecutive $K$ frames to enhance the robustness of compression.

The advantages of the embedding process can be summarized as follows:

(1) Try not to select the lowest frequency coefficients for robustness.
Usually, the method of modifying the frequency domain coefficients for embedding the watermark has to modify the coefficients with the lowest frequency band in order to improve the robustness of the algorithm. 
However, in this scheme, since multiple coefficients are selected for modification simultaneously, it is difficult for common attack methods to affect all modified coefficients.
So its robustness is better than that of only modifying a single coefficient.

(2) Since multiple coefficients of the same frequency band are modified in a single block, a low frequency band slightly close to the intermediate frequency can be selected for modification.
This approach can effectively reduce the impact of watermarks on image quality. 
And a good balance has been accessed between robustness and image quality.

\subsection{Watermarking extraction}
Watermark extraction is the inverse process of watermark embedding. 
The key point is to first find the block in which the watermark is embedded, and then judge whether the embedded information is +1 or -1 according to the rules of frequency band modification. 
As we have mentioned in Section 3.1, the blocks we choose are regions that are robust to microscopic coding attacks as well as macroscopic recording and geometric attacks. 
Therefore, for the video to be detected, the principle of block selection is still calculated by Equation~\ref{eq:total B}. 
The complete watermark detection process is as follows:

(1) The video to be detected is decoded and extracted to frames. Then frames will be transferred to the RGB color space.

(2) Select blocks according to Equation~\ref{eq:total B}, and mark these blocks as blocks to be detected.

(3) DWT and DCT are performed on the blue and green components of each block. And the coefficient sums of $cB$ and $cG$ are calculated according to the frequency band selected during embedding.

(4) Calculate the watermark information according to Equation~\ref{eq:extraction}.

\begin{equation}
    \label{eq:extraction}
    w(i)=\left\{
        \begin{aligned}
        &1& cB<cG \\
        &-1  &cB\geq cG
        \end{aligned}
    \right.
\end{equation}

\section{Experiments}

We choose 5 videos for evaluation in which 3 videos(News, Foreman and Mobile) are chosen from\cite{fitzek2004video}. These 3 videos are used by many watermarking methods for evaluation. All of which are in the format of YUV. Besides, we choose 2 another videos from Tencent Video library. We apply suitable transcoding  process to the 2 videos. Table~\ref{tab:videoinfo} gives more details of the 5 videos. \par

\begin{table}[htbp]
  \caption{Details of test videos}
  \label{tab:videoinfo}
  \begin{tabular}{lccccc}
    \toprule
    Video & Size(pixels) & FPS(fps) & Format & Length \\
    \midrule
    News & 352x288 & 25 & 4:2:0 & 12s \\
    Foreman & 352x288 & 25 & 4:2:0 & 12s \\
    Mobile & 352x288 & 25 & 4:2:0 & 12s \\
    Video1 & 1920x1080 & 25 & 4:2:0 & 30s \\
    Video2 & 1280x720 & 25 & 4:2:0 & 30s \\
    \bottomrule
  \end{tabular}
\end{table}

To demonstrate the effectiveness of our method, two types of attacks are applied to videos with watermark message.

Type I: Signal processing attack \par
\begin{enumerate}
    \item[$\bullet$] \textbf{rotate:} scale to suitable ratio and rotate to ensure no pixels are cropped
    \item[$\bullet$] \textbf{crop:} 4 sides are all cropped with 20\% ratio
    \item[$\bullet$] \textbf{resize:} both width and height are resized with a fixed ratio
    \item[$\bullet$] \textbf{projection:} scale and perspective projection
    \item[$\bullet$] \textbf{TLPF:} temporal low-pass filtering using consecutive four frames
    \item[$\bullet$] \textbf{FRC:} frame rate conversion,watermark extraction is done after converting the frame-rate back to 25 fps
\end{enumerate}

Type II: Recording attack \par
\begin{enumerate}
    \item[$\bullet$] \textbf{Recording:} use a digital device to capture videos played in a display.
\end{enumerate}
Compared to type 1, type 2 is a heavy attack and it will cause serious disturbance to the video information.

\subsection{Evaluation metrics}

\subsubsection{Quality measures} 

The perceptual quality of watermarked image is measured by Peak Signal to Noise Ratio (PSNR)\cite{lee2000high} and the mathematical equation of PSNR is given in below:

\begin{equation}
PSNR=10lg \frac {{255}^2} {MSE} 
\end{equation}

In above equation, MSE is defined as mean square error and given by \par

\begin{equation}
    MSE=\frac{1}{M\times N}\sum\limits_1^M \sum\limits_1^N (I(x,y) - I^*(x,y))^2
\end{equation}

In above equation, $I$ and $I^{*}$ is original host image and watermarked image respectively. \par

The MSE is measured in general scale while PSNR is measured in logarithmic scale. The high value of PSNR is indicated more imperceptibility of watermarking scheme.

\subsubsection{Robustness measures}
The robustness for any watermarking system is a very important requirement. To verify it, we apply to the watermarked video various types of attacks and we use the Bit Error Rate (BER) to compare the similarities between the original watermark w and the extracted watermark w*. The BER is calculated as:

\begin{equation}
BER=\frac{1}{P}\sum\limits_{j=1}^P {\mid {w^*(j) - w(j)} \mid}
\end{equation}

where $w$, $w^{*}$, and $P$ are, respectively, the original watermark, the extracted watermark, and the size of the watermark.

\subsection{Experimental results}
Table~\ref{tab:psnrinfo} shows the average PSNR of the watermarking videos and the increase in bit rate.
Since our method is embedded in the RGB color space, while other methods are embedded in the YUV color space, to ensure an objective and fair comparison, we calculate the average value of the PSNR of 3 channels.
It is easy to see that our method achieves the best results in PSNR, which means that our method has minimal impact on video quality. 

The reason we can embed the watermark almost losslessly is that we choose to embed the watermark only on the blue component. 
Since the Y channel concentrates all the brightness information of the frame, the modification on the Y channel will cause all three components of RGB to be affected, thereby reducing the value of PSNR.
In addition, instead of blindly pursuing low-frequency embedding, our method can better reduce the effect of watermarking on video quality by selecting mid-low frequency bands.

\begin{table*}[htbp]
\centering
  \caption{Average PSNR and bit rate increase due to watermarking}
  \label{tab:psnrinfo}
  \begin{tabular}{lccccccc}
    \toprule
    Video & PSNR (ours) & PSNR (Wang\cite{wang2006blind2}) & PSNR (Celik\cite{celik2007camcorder}) & PSNR (Tancik\cite{tancik2020stegastamp}) & $q_s$ = 4(\%) & $q_s$ = 8(\%) & $q_s$= 12(\%) \\
    \midrule
    News & 53.96 & 48.12 & 47.24 & 42.73 & 6.73 & 5.52 & 4.27 \\
    Foreman & 51.37 & 47.09 & 45.88 & 40.57 & 4.18 & 4.09 & 3.64 \\
    Mobile & 49.68 & 46.35 & 46.31 & 40.11 & 6.05 & 5.66 & 4.08 \\
    Video1 & 50.83 & 46.93 & 46.57 & 41.83 & 5.47 & 4.89 & 3.71 \\
    Video2 & 49.25 & 45.27 & 46.11 & 40.25 & 5.93 & 5.31 & 4.22 \\
    \bottomrule
  \end{tabular}
\end{table*}

In addition to the impact on video quality, the robustness of the watermark is also an aspect that we are concerned about.
Table~\ref{tab:berinfo} shows the bit error rate of watermarking under various attacks, including geometric attack, timing attack, and camcorder recording attack.
It can be concluded from Table~\ref{tab:geo attack} that our method can ensure the watermark information not being destroyed to the greatest extent in geometric attacks such as rotation, scaling, and projection transformation. 
The bit error rate of the extracted information is only about 1\% after being geometrically attacked. 
What's more, under the cropping attack, our method can extract the complete watermark information without the loss of watermark bit information, while other methods achieve high bit error rates, especially Wang's method even up tp 40\%. 
This is due to the fact that we only selected the middle area of the image to construct the block during embedding.

In addition to the geometric attack, it can be seen from the last three rows of Table ~\ref{tab:geo attack} that the watermarked videos are also subjected to a timing attack. 
The result is still significantly higher than the effect of the other three groups. 
From the principle of embedding, the method of embedding one-bit information in consecutive $K$ frames can prevent the watermark information from being damaged by the influence of adjacent frames during video coding. 
In addition, when extracting the watermark, the frequency domain coefficients are calculated after calculating the average frame by using consecutive $K$ frames.
This method of taking the mean value effectively concentrates the medium and low frequency information of consecutive frames, which can better extract the watermark information from the frequency domain coefficients.

\begin{table}[htbp]
  \caption{The average bit error rate under several attacks}
  \label{tab:berinfo}
  
  \begin{subtable}{0.45\textwidth}
  
      \begin{tabular}{lcccc}
        \toprule
        Attack & Proposed(\%) & Wang(\%) & Celik(\%) & Tancik(\%) \\
        \midrule
        Rotate 4°        & 1.5 & 10.2 & 22.6 & 5.6 \\
        Crop 20\%       & 0  & 40+ & 2.7 & 11.8 \\
        Resize x 1.5    & 0.4 & 5.1 & 6.7 & 3.3 \\
        Resize x 0.5    & 0.7 & 5.0 & 5.3 & 3.7 \\
        Projective      & 1.2 & 7.9 & 21.3 & 4.9 \\
        TLPF            & 3.2 & 28.2 & 9.3 & 10.4 \\
        FRC 30fps       & 1.1 & 40+ & 2.7 & 5.5 \\
        FRC 20fps       & 1.9 & 40+ & 2.7 & 7.1 \\
        \bottomrule
      \end{tabular}
      \caption{ Signal processing attack\label{tab:geo attack}}
  \end{subtable}
  
  \begin{subtable}{0.45\textwidth}
  
      \begin{tabular}{lcccc}
        \toprule
        Attack & Proposed(\%) & Wang(\%) & Celik(\%) & Tancik(\%) \\
        \midrule
        recording 1      & 1.7 & 40+ & 24.5 & 6.3 \\
        recording 2      & 2.1 & 40+ & 27.5 & 4.9 \\
        \bottomrule
      \end{tabular}
      \caption{ Low-definition video recording attack\label{tab:camcorder1}}
  \end{subtable}
  
  \begin{subtable}{0.45\textwidth}
      
      \begin{tabular}{lcccc}
        \toprule
        Attack & Proposed(\%) & Wang(\%) & Celik(\%) & tancik(\%) \\
        \midrule
        recording 1      & 2.2 & 40+ & 21.3 & 5.7 \\
        recording 2      & 1.3 & 40+ & 26.7 & 6.0 \\
        \bottomrule
      \end{tabular}
      \caption{ High-definition video recording attack\label{tab:camcorder2}}
  \end{subtable}
  
\end{table}
 
\begin{figure} [!h]
	\centering
	\subfloat[\label{fig:a}]{
		\includegraphics[scale=0.055]{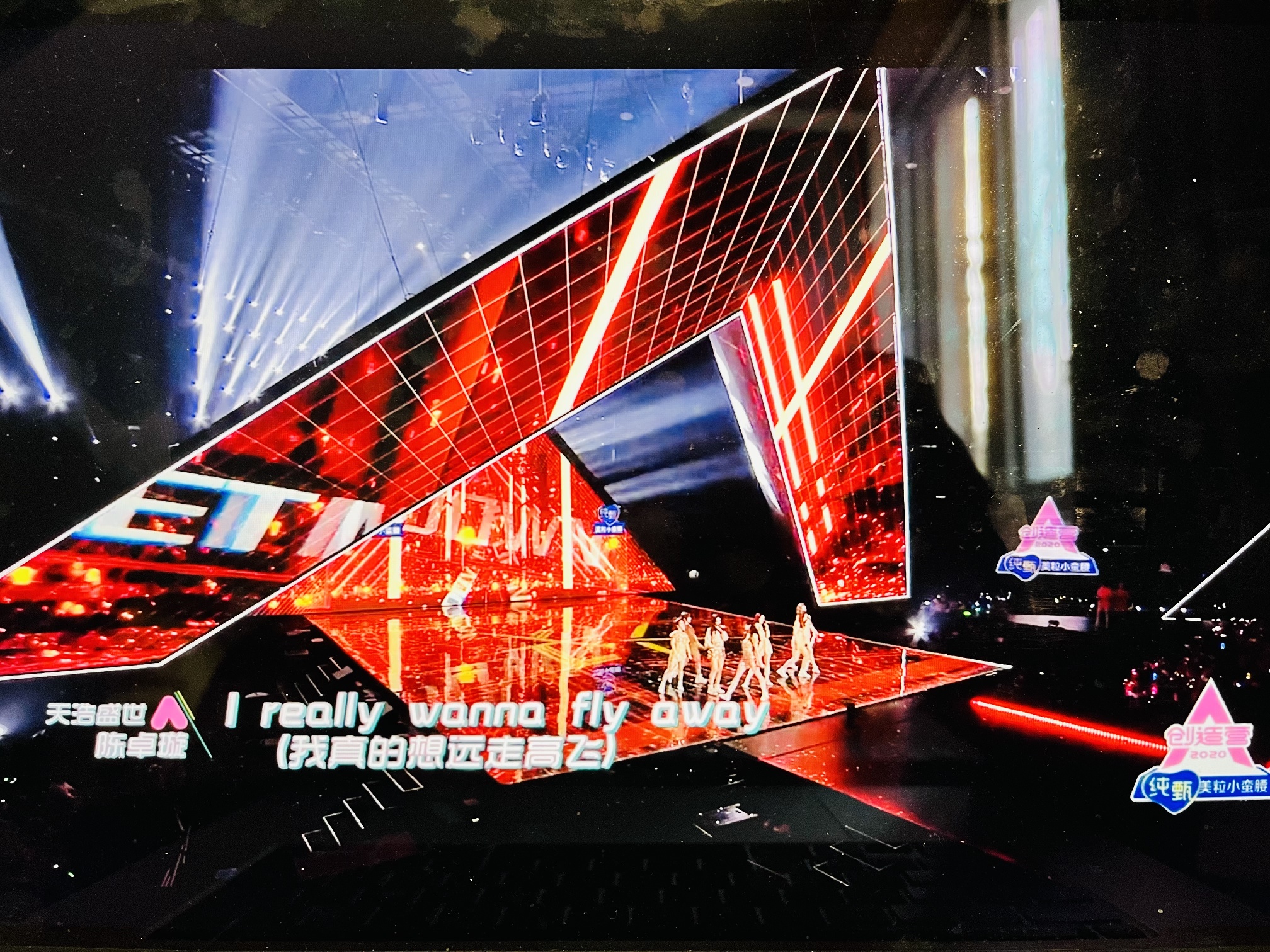}}
	\subfloat[\label{fig:c}]{
		\includegraphics[scale=0.055]{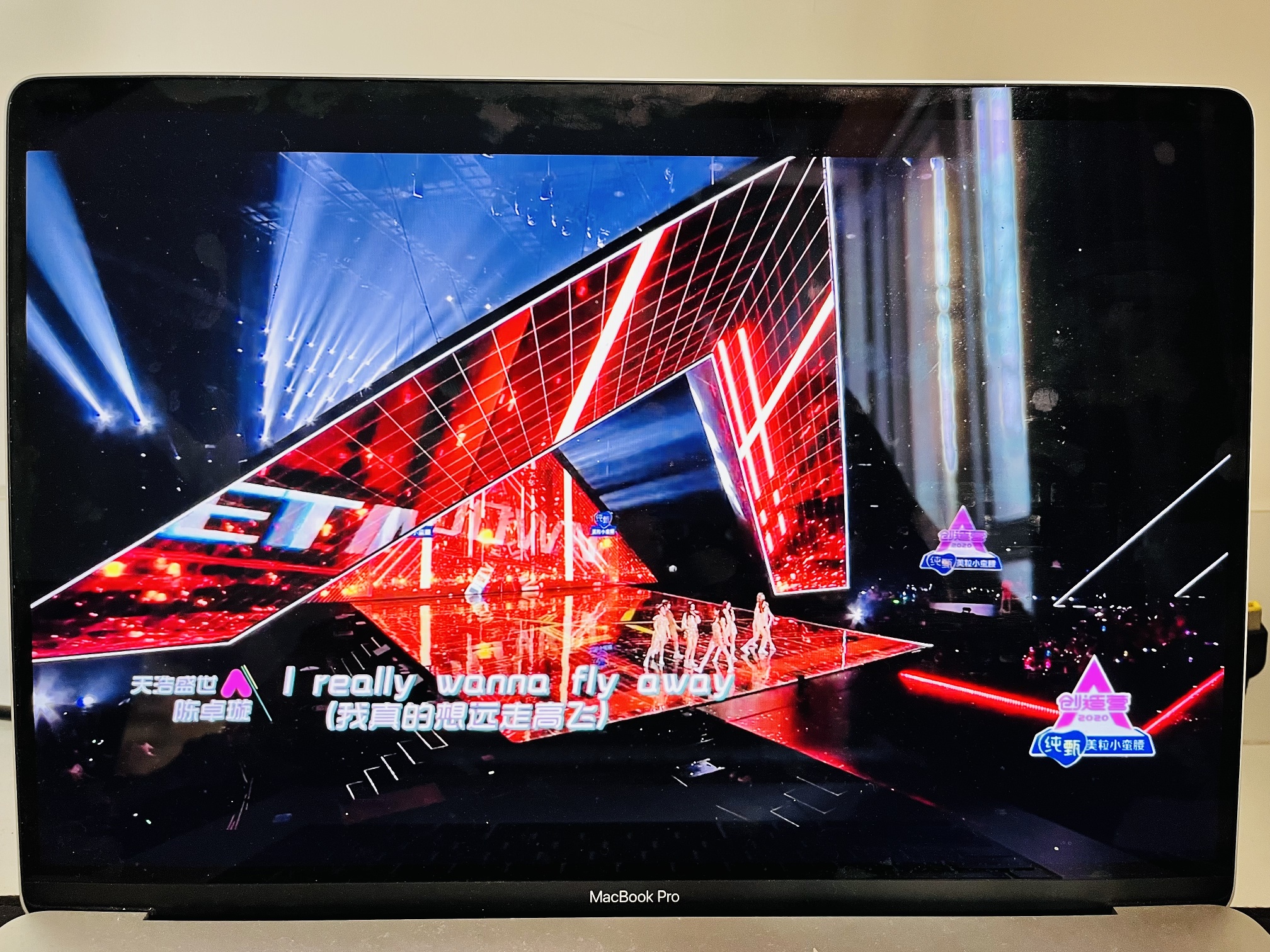}}
	\\
	\subfloat[\label{fig:b}]{
		\includegraphics[scale=0.055]{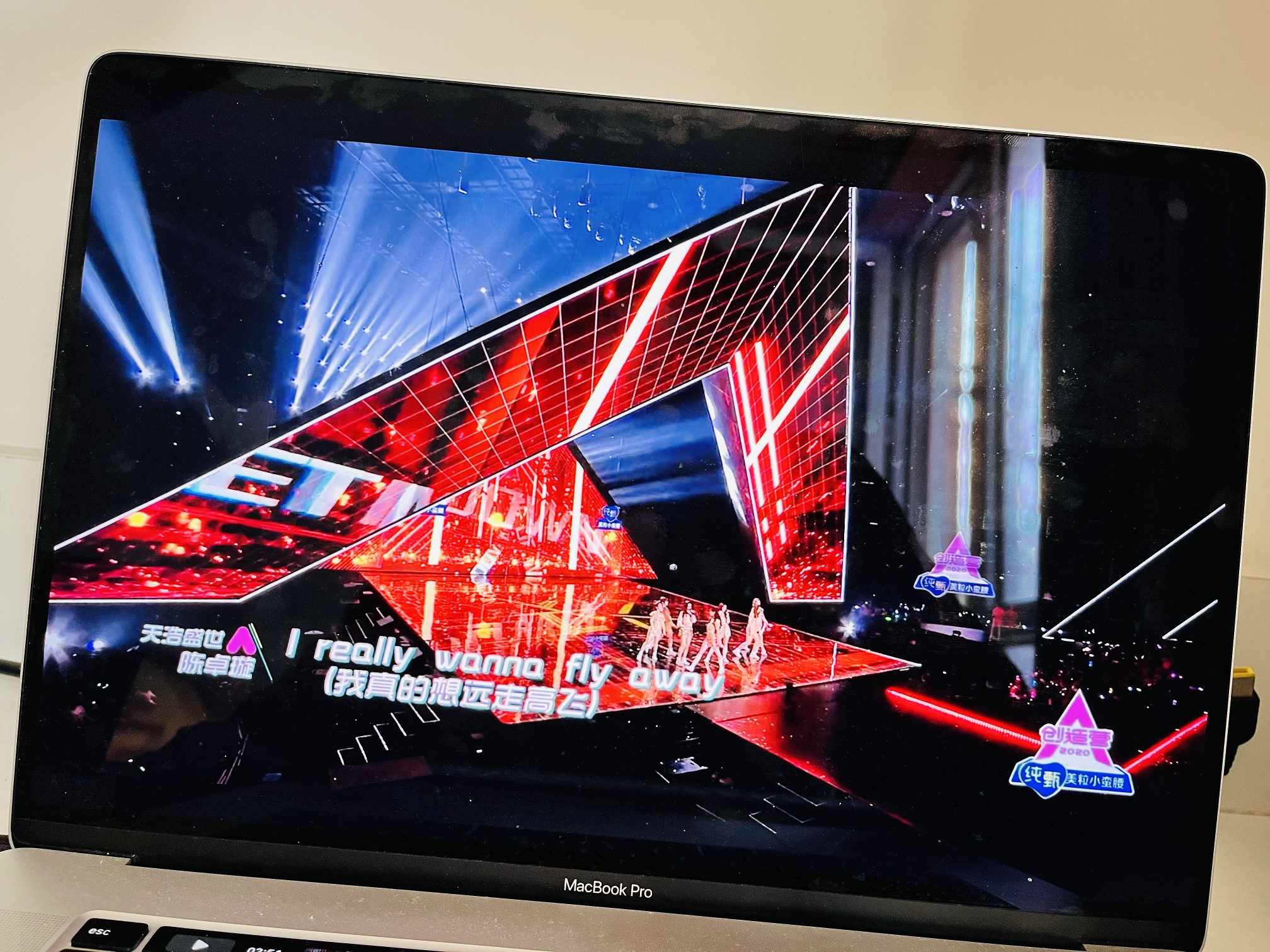} }
	\subfloat[\label{fig:d}]{
		\includegraphics[scale=0.055]{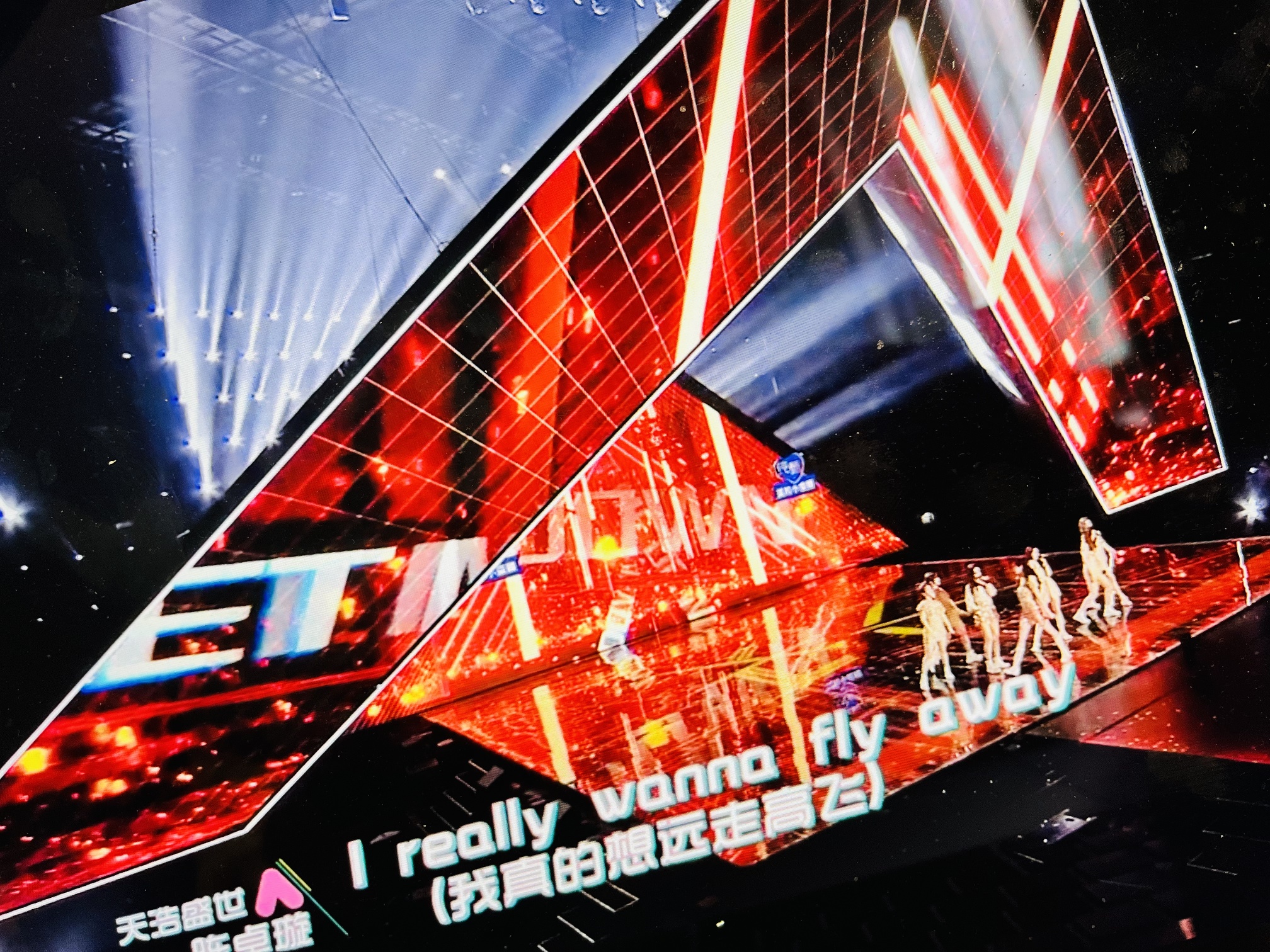}}
	\caption{These four figures show the effect of camcorder recording in our experiments. All videos are shooting by iPhone 13 using the wide-angle lens. Fig (a) shows an ideal case of camcorder recording. The content in the screen has not been geometrically deformed, and the content of the external environment has not been introduced. Fig (b) represents the camcorder recording effect with only redundant content introduced. Fig (c) and (d) represent strong camcorder recording attacks. The content of Fig (c) is attacked by a serious perspective transformation, and the external environment content is also added, while Fig (d) is under perspective transform attacks involving cropping }
	\label{fig:camcorder} 
\end{figure}

Table~\ref{tab:camcorder1} and Table~\ref{tab:camcorder2} show the detection results of watermark extraction under camcorder recording. 
In these experiments, several recordings of each video are made by iPhone 13 (1080p, 30FPS) with 1.0m away handheld shooting from a 16 inch retina display screen (Macbook Pro).
The frame rate of the recorded video is converted back to the original 25 fps and the video content is aligned for watermark extraction. 
Camcorder recording introduces attacks of rotation, panning, zooming, brightness changes, noise, and slight shaking caused by hand-holding. 
Fig~\ref{fig:camcorder} shows four examples of camcorder recordings.

In Table ~\ref{tab:camcorder1}, we compare the watermark detection results after camcorder recording on \emph{News}, \emph{Foreman}. 
It can be seen that our method achieves quite good results with bit error rates of only 1.7\% and 2.1\%. 
In addition, we also select some high-resolution movie clips to verify the experimental results. 
Recording 1 and Recording 2 in Table ~\ref{tab:camcorder2} are two high-definition video clips with duration of 60s, resolution of 1080p and bit rate of 5M. 
For each video, we apply four different levels of attack as shown in Fig~\ref{fig:camcorder} and calculated the average bit error rate for these four levels.
Obviously our method also achieves the best results on these videos.

Considering that the camcorder recording attack changes the frame rate of the original video, when extracting the watermark, we first convert the video frame rate to the original frame rate. 
Then, the average value of the RGB components of consecutive K frames is calculated separately. 
And the watermark detection is conducted on this average frame.

Camcorder recording attack is a very complex combined attack. 
Besides frame rate changes, camcorder recording can be regarded as a combination of geometric attacks (rotation, translation, scaling) and pixel value attacks (moire, illumination changes, etc.). 
Specifically, since the watermarking blocks we choose are very robust and insensitive to geometric attacks, displacement and rigid deformation caused by geometric attacks do not destroy these blocks.
For changes in pixel values such as moire, illumination changes, etc., it usually does not change the relative difference of the blue and green components, so the watermark information can be well preserved under these attacks. 
In addition, since we disperse the modification amount during watermarking on multiple frequency domain coefficients, some tiny noise points introduced by the camera screen will not affect all the modified coefficients at the same time, which makes our method more robust to camcorder recording attacks.

To further verify the effect of our algorithm, ablation experiments are also conducted as shown in Table~\ref{tab:ablation}. 
We compare the effect of not using texture factor and not using ORB feature. 
It is obviously that under video compression, geometric attack and recording attack, the watermark detection results of these two groups are not as high as the results of using texture factor and ORB feature simultaneously. 
And without filtering out some blocks by taking intersection, the number of blocks selected by using only texture factor and only ORB feature will be more than using both at the same time.
The more blocks there are, the more areas of the image will be modified, and the more video quality will be affected.
Therefore, the block selection method we designed can not only improve the robustness of the watermarking algorithm, but also reduce the impact of watermark embedding on the video quality.

\begin{table}[htbp]
\centering
  \caption{Influence of block selection method on watermark effect}
  \label{tab:ablation}
  \begin{tabular}{cccccc}
    \toprule
    Texture factor & ORB & \makecell{Rotation \\ (4°)} & \makecell{Resize \\ (x 0.5)} & Recording & PSNR \\
    \midrule
    \Checkmark & \XSolid & 5.3\% & 4.6\% & 11.9\%  & 47.61\\
    \XSolid & \Checkmark & 3.6\% & 3.1\% & 8.2\% & 49.50\\
    \Checkmark & \Checkmark & 1.5\% & 0.6\% & 2.1\%  & 51.37\\
    \bottomrule
  \end{tabular}
\end{table}


\section{Conclusion}
This paper proposes a highly robust watermarking algorithm based on adaptive region selection and channel reference.
By analyzing the content information of the video including texture information and feature points, areas that are not easily disturbed are selected to add watermarks.
During the watermark embedding process, we choose to modify the frequency domain coefficients of the blue component instead of the luminance component.
This approach refers to the principle of HVS, which effectively reduces the impact of watermarks on image quality. 
Experiments show that our method is not only highly robust to geometric attacks, but also has very good resistance to timing attacks and camcorder recordings.
Our future work will focus on more practical scenarios, such as shortening the time required for watermark extraction, adapting to different terminal devices, and supporting HDR materials, etc.
\bibliographystyle{ACM-Reference-Format}
\bibliography{main}

\end{document}